\documentclass[amsmath,amssymb,aps,pra,twocolumn]{revtex4-2}
\usepackage{physics}
\usepackage{xcolor}
\usepackage{graphicx}
\usepackage{amsmath}
\usepackage{amssymb}
\usepackage{bbold}
\usepackage{bm}
\usepackage{mathtools}

\newcommand{\cor}[1]{\textcolor{black}{#1}}
\newcommand{\corb}[1]{\textcolor{black}{#1}}

\begin{document}
\title{Optimal estimation of conjugate shifts in position and momentum by classically correlated probes and measurements}
\begin{abstract}
\cor{
Multi-parameter estimation is necessary for force sensing due to simultaneous and nontrivial small changes of position and momentum. Designing quantum probes that allow simultaneous estimation of all parameters is therefore an important task. The optimal methods for estimation of the conjugate changes of position and momentum of quantum harmonic oscillator employ probes in entangled or quantum non-Gaussian states. We show that the same results can be obtained in a significantly more feasible fashion by employing independent sets of differently squeezed Gaussian states classically correlated to position or momentum measurements. This result demonstrates an unexplored power of a classical correlation between the probe states and measurements directly applicable to force sensing
}
%
\end{abstract}

\author{Kimin Park}
\affiliation{Department of Optics, Palack\'y University, 17. listopadu 1192/12, 771 46 Olomouc, Czech Republic}
\author{Changhun Oh}
\affiliation{Department of Physics and Astronomy, Seoul National University, Seoul 08826, Korea}
\affiliation{Pritzker School of Molecular Engineering, The University of Chicago, Chicago, Illinois 60637, USA}
\author{Radim Filip}
\affiliation{Department of Optics, Palack\'y University, 17. listopadu 1192/12, 771 46 Olomouc, Czech Republic}
\author{Petr Marek}
\affiliation{Department of Optics, Palack\'y University, 17. listopadu 1192/12, 771 46 Olomouc, Czech Republic}
\maketitle

\section{Introduction}
\cor{The main goal of quantum metrology lies in finding and achieving the ultimate limits on measuring parameters of known physical processes \cite{Giovannetti11,Demkowicz15,Holevo,polino2020}. It can be applied to estimation of single \cite{Giovannetti11,Demkowicz15,RevModPhys.90.035005,Paris_09, pirandola2018,RevModPhys.89.035002} as well as multiple parameters
\cite{humphreys2013,vidrighin2014, PhysRevLett.121.130503, demkowiczdobrzaski2020,liu2019,PhysRevLett.120.080501,PhysRevLett.106.090401,PhysRevLett.117.190802,PhysRevA.96.062107,PhysRevA.98.012114}. The key principle lies in paying attention to the quantum state of the employed probes such as modes of light \cite{Mitchell04,Aasi_13}, mechanical modes of trapped ions \cite{burd2019,mccormick2019,wolf2019motional}, collective modes of spins in a magnetic field \cite{PhysRevLett.116.030801,PhysRevLett.119.043603,PhysRevLett.127.193601}, mechanical modes of optomechanical oscillators \cite{qvarfort2018gravimetry,PhysRevLett.125.213601,fogliano2021}, or in principle any other quantum system.
}

\cor{Quantum multi-parameter estimation aims to discern a set of real parameters $\theta = (\theta_1,\cdots,\theta_M)$ that characterize a given channel. This is realized by preparing an ensemble of probes in a known quantum state, sending them through the channel, subjecting them to quantum measurement, and subsequently applying an estimation strategy to produce a set of unbiased estimators $\tilde{\theta} = (\tilde{\theta}_1,\cdots,\tilde{\theta}_M)$. The quality of the estimation depends on the difference between the estimators and the true values and it can be conservatively evaluated by the mean quadratic variances $v_j := \langle (\theta_j - \tilde{\theta}_j)^2\rangle$ \cite{Holevo}. For any particular measurement strategy, these variances are bounded by the inverse of the classical Fisher information (CFI) Their ultimate limit is given by the Holevo-Cramer-Rao (HCR) bound \cite{Holevo,demkowiczdobrzaski2020} that can be obtained by minimization over all possible measurement strategies, which can, in many cases, be only done numerically. Numerical computation can be also used to obtain the Nagaoka-Hayashi bound for separable single copy measurements \cite{nagaoka2005,conlon2021}. The variances are also lower bounded by the inverse of the quantum Fisher information (QFI) obtained either from symmetric or right logarithmic derivative \cite{Holevo, Paris_09,liu2019}, but this bound is not always tight for multi-parameter quantum estimation. The optimal probe state is such that offers the maximal precision for constraints limiting probe preparation, sampling, and measurement. Some commonly employed constraints are preparing the probe states as identical copies of a specific quantum state \cite{Giovannetti11,demkowiczdobrzaski2020,gill2013}, and maximal energy of the probes \cite{PhysRevA.87.012107,PhysRevA.95.012305,PhysRevA.97.012106,wolf2019motional}.
}

\cor{
One essential task of quantum sensing is estimation of parameters of a small mechanical, electrical, magnetic, or optical force \cite{poggio2008,PhysRevA.88.042112,PhysRevA.94.052115}.
A particular scenario commonly studied in this context is the simultaneous estimation of the position and momentum - two parameters of quantum displacement acting on a state of harmonic oscillator  \cite{PhysRevA.87.012107,PhysRevA.97.012106, PhysRevResearch.2.023182, PhysRevA.95.012305}. Beyond the considerable fundamental interest \cite{PhysRevA.98.012114,demkowiczdobrzaski2020}, this basic measurement is already relevant for calibration of continuous variables quantum key distribution in optical systems \cite{PhysRevA.97.032329}, estimation of weak electric fields with trapped ion crystals \cite{gilmore2021}, or estimation of temperature in ultracold lattice gasses \cite{mehboudi2015}.
On the elementary level, the displacement is represented by unitary evolution operator $D(\mu+i\nu) = e^{-i \nu X -i \mu P}$, where $X$ and $P$ are the quadrature operators of the optical field with commutator $[X,P] = i$ and $\mu$ and $\nu$ are the unknown parameters. This scenario is an example of a Gaussian shift model \cite{demkowiczdobrzaski2020} and the difficulty lies in simultaneously estimating values of non-commuting operators $X$ and $P$ bound together by Heisenberg uncertainty relations.}

\cor{
There have been two approaches suggested to overcome this issue. The first approach utilizes a set of identical probes in two mode squeezed quantum entangled states in which both quadratures can be simultaneously estimated with high precision due to the dense coding effect \cite{PhysRevA.87.012107}. The second approach employs a set of identical single mode probes prepared in quantum states that are superpositions of differently displaced squeezed states \cite{PhysRevA.95.012305,PhysRevResearch.2.023182}. In this case the possibility of simultaneous estimation is the consequence of the rich sub-Planck structure of the non-classical and highly non-Gaussian quantum states. A similar technique was also suggested for estimation of amplitude of the displacement operation, which is only a single parameter but depends on the two non-commuting quantities $X$ and $P$. The amplitude is efficiently estimated by employing probes in Fock states of the harmonic oscillator \cite{wolf2019motional,Oh_2020}, which are again non-Gaussian states with rich sub-Planck structure. These approaches are argued to be optimal in the sense that the proposed measurement strategies saturate the HCR bound of the probe states and that the probe states have the minimal energy that allows this HCR \cite{PhysRevA.87.012107,PhysRevA.95.012305,PhysRevA.97.012106,wolf2019motional}.
}

\cor{In this paper we show that the optimal estimation of the two parameters of displacement can be, on average and in the limit of large number of probes, achieved also by measuring the two parameters independently by using two sets of factorized quantum states, squeezed in position or momentum, and measurement of the respective variable. This approach, inspired by some techniques of quantum process tomography \cite{Rahimi_Keshari2011, Safranek2016, Teo2021}, leads to the same mean quadratic error as the methods based on entangled two-mode squeezed states \cite{PhysRevA.87.012107} or quantum non-Gaussian states \cite{PhysRevA.95.012305,PhysRevResearch.2.023182} with the same energy of the individual probes. This performance is obtained even though the sequence of squeezed states is generally more feasible to implement.}


\section{Methods}
\begin{figure}
\centering
\includegraphics[width=7.5cm]{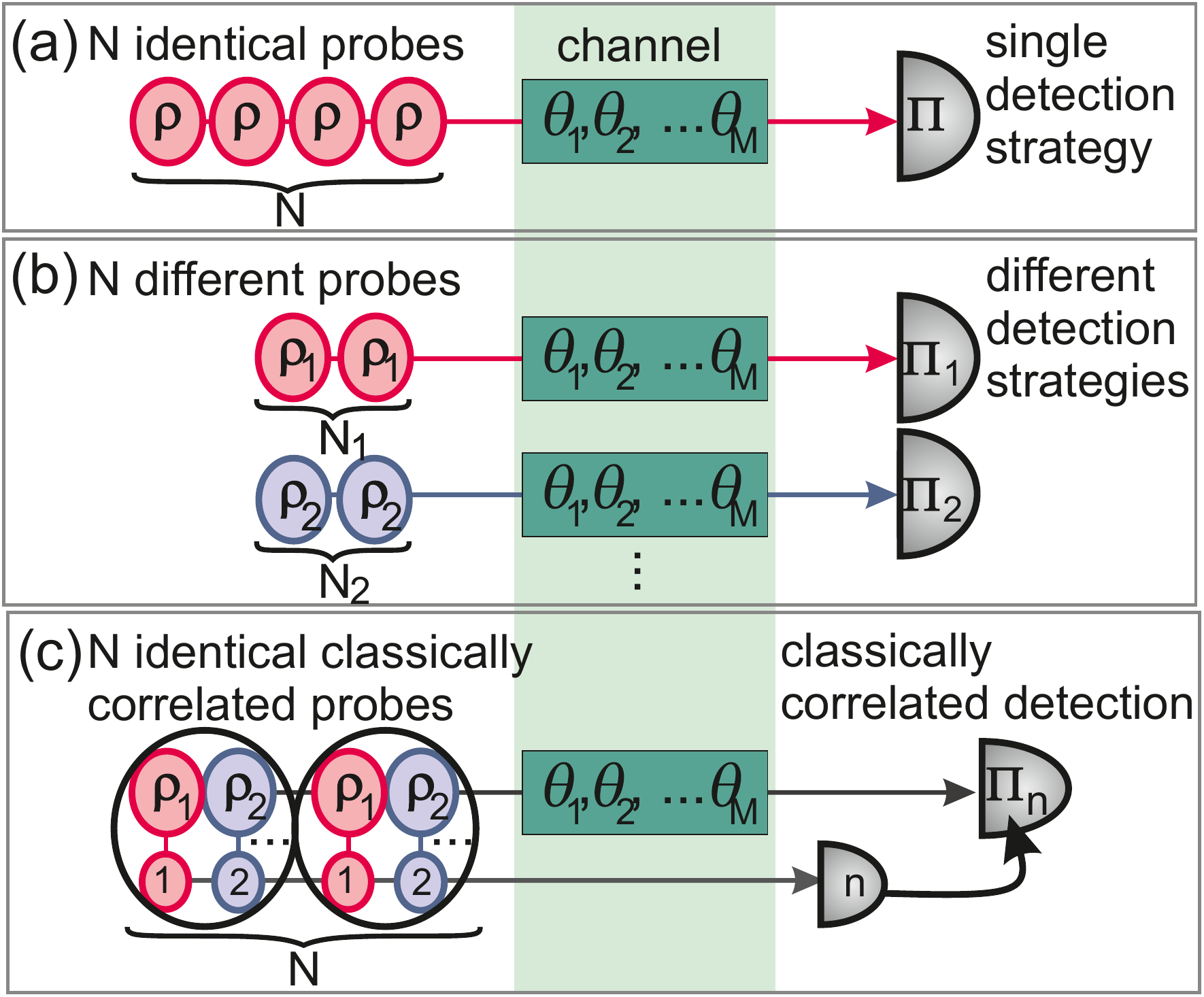}
\caption{\cor{a) Standard scenario of quantum parameter estimation with a set of identical probes $\rho$ which interact with the channel, undergo transformation given by parameters $\theta_1,\cdots, \theta_M$, and are measured by measurements given by POVM $\Pi$. b) Quantum parameter estimation by different separate probes $\rho_1, \rho_2, \cdots$ that individually interact with the channel and are individually measured by fixed measurements with POVMs $\Pi_1, \Pi_2, \cdots$ classically correlated with the respective probes. c) Quantum parameter estimation by different separate probes, such as those in Fig.~\ref{schl}b, effectively represented by estimation with a set of identical classically correlated states (\ref{rhot}) of probes $\rho_1,\rho_2\cdots$ and their respective orthonormal markers denoted by $n = 1,2, \cdots$. The classically correlated measurement (\ref{povms}) consists of detection of the marker state $|n\rangle$ followed by feed-forward setting the particular detector to the one with POVM $\Pi_n$.}}
\label{schl}
\end{figure}

The standard estimation scheme with a single set of identical probes represented by quantum state $\rho^{\otimes N}$ and a single measurement strategy given by positive operator valued measure (POVM) $\Pi$ is illustrated in Fig.~\ref{schl}a. Each probe independently interacts with the channel and is subsequently measured. Note that each individual probe can be in an entangled state of several subsystems of which only some interact with the channel. The protocol we propose is illustrated in Fig.~\ref{schl}b and it employs several sets of different probes, each one prepared in one of different states $\rho_n$. Their collective state can be therefore represented by $\bigotimes_{n}\rho_n^{\otimes N_n}$ where the numbers of probes satisfy $\sum_n N_n = N$. Each probe interacts with the channel that transforms its state into $\rho_n(\theta)$  and is measured by a measurement $\Pi_n$ tailored to the respective set. Each such measurement is represented by POVM elements  $\{\Pi_n(o_n)\}$ with the corresponding measurement results $\{o_n\}$. We do not assume specific dimensionality of the measurement, the measurement results $o_n$ are vectors of real values. The estimators $\tilde{\theta}$ can now be obtained from the classical probability distributions $P(o_n)$.

\cor{For the sake of straightforward comparison we can also express the protocol with sets different probes in an effective form that utilizes a single set of identical classically correlated probes, see Fig.~\ref{schl}c. Note that this is only for the sake of comparison and clarity, in practical application it is not necessary. In this approach, each individual probe state can be expressed as a single classically correlated quantum state }
\begin{equation}\label{rhot}
    \rho_{T} = \sum_n w_n |n\rangle\langle n| \otimes \rho_n,
\end{equation}
\cor{ where $w_n = N_n/N$ are the relative probe weights and $|n\rangle$ are arbitrary orthonormal states.} Their role is to serve as classical markers \cor{for the detectors}, differentiating the individual probe states in the mixture. 
The probe is then subjected to a trace preserving quantum channel which transforms the state into $\rho_{T}' = \sum_{n=0} w_n |n\rangle\langle n| \otimes \rho_n'(\theta)$, leaving the marker states $|n\rangle$ as well as the weights $w_n$ unchanged.

The probe state $\rho_{T}'$ now needs to be measured to extract the information imparted by the channel. In this effective model, we consider this measurement to be composed of two parts. First we measure the marker states. Since they are orthogonal and not affected by the channel, the measurement always returns the correct marker, which will be used to set the desired measurement for the probe. Formally, the global POVM of the measurement is such that the measured values are vectors of real values $ [n,o_n]$ each of them corresponding to POVM element
\begin{equation}\label{povms}
    |n\rangle\langle n| \otimes \Pi_n(o_n).
\end{equation}
The joint probability distribution obtained by this measurement is then represented by probability distribution
\begin{equation}\label{}
    f(n,o_n) = \mathrm{Tr}[\rho_T' |n\rangle\langle n| \otimes \Pi_n(o_n)]
\end{equation}
and it is essentially a finite sequence of probability distributions $f(o_n)$ for each individual combination of the probe state and its respective measurement.
The joint probability distribution can be used to evaluate elements of the classical Fisher information (CFI) matrix $\mathbf{C}_T$ for any pair of channel parameters $\theta_j$, $\theta_l$:
\begin{equation}\label{cfi}
    \mathbf{C}_T(j,l) = \sum_{n=1}  \int d o_n \frac{ (\partial_{\theta_j}f(n,o_n) \partial_{\theta_l} f(n,o_n))}{f(n,o_n)},
\end{equation}
where the integration always goes over the full support of variable vector $o_n$ and the sum is over all the marker states in (\ref{rhot}). Since $f(n,o_n) = w_n f_n(o_n)$, where $f_n(o_n) = \mathrm{Tr}[\rho_n \Pi(o_n)]$ is the probability distribution for the specific combination of probe and measurement, we can see that the classical Fisher information  matrix obtained in this way is equal to the weighted sum of CFI matrices of the individual probe-measurement pairs with weights $w_n$,
$    \mathbf{C}_T  = \sum_n w_n \mathbf{C}_n.$
%
\cor{Obtaining the tight HCR bound is not as straightforward and usually needs to be performed numerically \cite{PhysRevLett.123.200503}. In the other hand, the non-tight Cramer-Rao bound based on quantum Fisher information (QFI) matrix can be easily obtained from $\mathbf{Q}_T = \sum_n w_n \mathbf{Q}_n$, which follows from the additivity property. Note that in the following we will be interested in the comparison of a specific estimation scenario for which the CFI will be sufficient. }

\begin{figure}
\centering
\includegraphics[width=7cm]{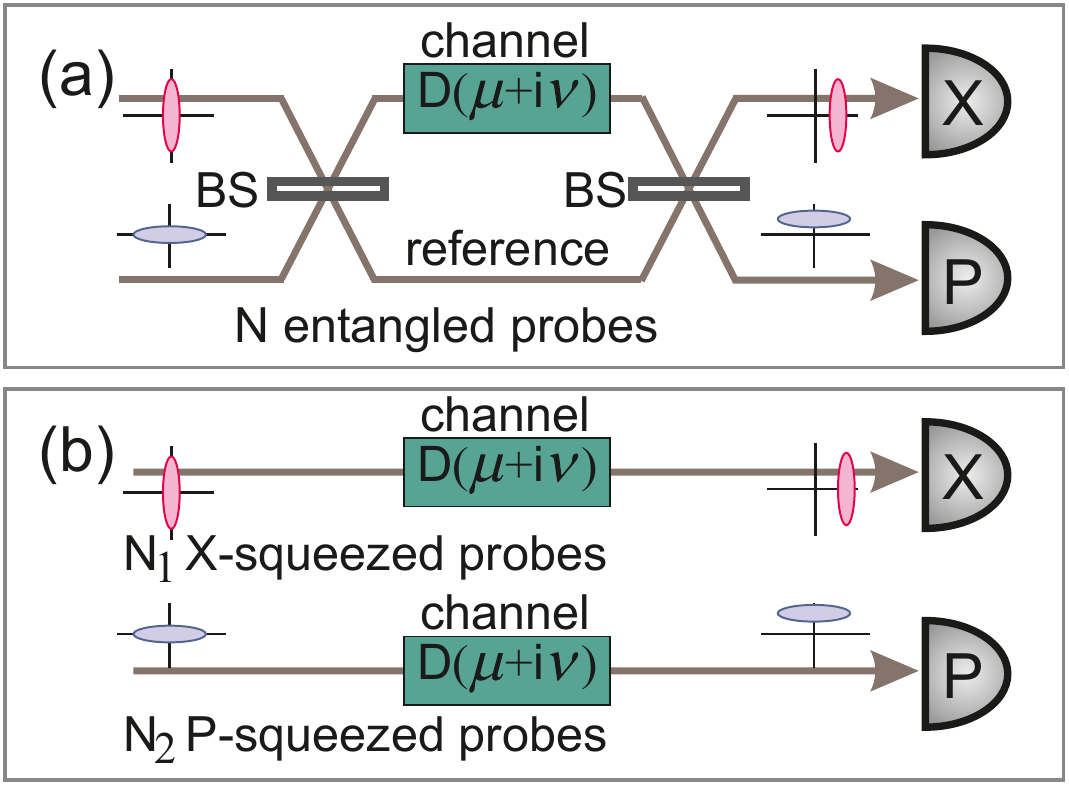}
\caption{ \cor{Estimation of two parameters of displacement operation $D(\mu + i \nu) = e^{-i \nu X -i \mu P}$. a) Entanglement based scenario in which each of the $N$ probes is prepared in an entangled state by interfering two orthogonally squeezed states on a balanced beam splitter (BS). One of the modes passes through the estimated channel, while the second serves as a reference. The modes are recombined on another balanced beam splitter and quadratures $X$ and $P$ are measured by homodyne detectors. b) Estimation by two sets of different probes of which $N_1$ is squeezed in $X$ and $N_2$ is squeezed in $P$, $N_1 + N_2 = N$. Each probe passes through the channel and the corresponding quadratures are measured by a homodyne detection. }}
\label{fig2}
\end{figure}

\section{Results}
Let us now turn to the specific case of estimating the two parameters of the displacement operation $D(\mu + i \nu) = e^{-i \nu X -i \mu P}$. The entanglement based protocol \cite{PhysRevA.87.012107,PhysRevA.97.012106, PhysRevResearch.2.023182} for simultaneous estimation of $\mu$ and $\nu$ is illustrated in Fig.~\ref{fig2}a. Each probe is an entangled two-mode squeezed state that can be prepared by interference of two orthogonally squeezed states on a balanced beam splitter. The state of the probe can be explicitly written in the $X$ representation as:
\begin{equation}\label{psient}
    |\psi\rangle = \frac{1}{ 2\pi }\int dx~ dy~ e^{-(x+y)^2/4 e^{-2r}} e^{-(x-y)^2/4 e^{2r}}|x,y\rangle.
\end{equation}
This is an entangled quantum state defined by non-classical correlations $\langle (X_1-X_2)^2\rangle = \langle (P_1 + P_2)^2\rangle = e^{-2r}$ of the quadrature operators of the respective two modes. The optimal measurement is composed of another balanced beam splitter followed by homodyne detectors measuring quadratures $X$ and $P$ of the two output ports, \cite{PhysRevA.97.012106}, that produce joint probability distribution
\begin{equation}\label{}
    f(x,p) = \frac{1}{\pi e^{-2r}} \exp[-\frac{(x-\mu/\sqrt{2})^2}{e^{-2r}} - \frac{ (p-\nu/\sqrt{2})^2}{e^{-2r}}].
\end{equation}
\cor{From here we can arrive at a diagonal CFI matrix with elements $C_E(1,1) = C_E(2,2) = e^{2r}$  }
\cor{that saturate the HCR bound \cite{PhysRevA.87.012107} for the case when the two displacement are given equal importance. When this is not the case, the protocol can be adjusted by changing the symmetry of the probe as well as of the final measurement. The HCR can be reached in all the cases \cite{PhysRevResearch.2.023182}. }

\cor{The particular application of method depicted in Fig.~\ref{schl}b is illustrated in Fig.~\ref{fig2}b. It is based on two sets of vacuum states squeezed either in the $X$ or the $P$ quadrature, which are measured by homodyne detection of the respective squeezed quadrature. We can formally write the effective mixed state (\ref{rhot}) as}
\begin{equation}\label{2sstate}
    \rho_T = w_1 |\phi\rangle\langle \phi|\otimes S |0\rangle\langle 0|S^{\dag} + w_2|\phi^{\bot}\rangle\langle \phi^{\bot}|\otimes S^{\dag} |0\rangle\langle 0|S,
\end{equation}
where $|\phi\rangle$ and $|\phi^{\bot}\rangle$ are the orthonormal marker states, $|0\rangle$ is the vacuum state of quantum harmonic oscillator, and $S$ is a squeezing operator such that $S|0\rangle$ is a squeezed vacuum state with variance $\mathrm{var}~X = \frac{1}{2}e^{-2r}$. Note that the squeezing was chosen in such a way that the probe states are the same as those used to compose the entangled state (\ref{psient}) in the entanglement based protocol. As a consequence, the same energy passes \cor{through the channel} during each run of the protocol and they are therefore directly comparable. The homodyne measurements aligned with the squeezed quadrature of the probe states are represented by POVM elements
\begin{align}\label{}
    \Pi(1,q) = |\phi\rangle\langle \phi|\otimes |X=q\rangle \langle X=q|, \nonumber \\
     \Pi(2,q) = |\phi^{\bot}\rangle\langle \phi^{\bot}| \otimes |P = q\rangle\langle P=q|,
\end{align}
where $|X=q\rangle$ and $|P = q\rangle$ represent the $X$ and $P$ quadrature eigenstates with eigenvalue $q$, respectively. \cor{These measurements produce a pair of probability distributions $f_X(x) = w_1 \exp(-(x-\mu)^2/e^{-2r})/\sqrt{\pi e^{-2r}}$ and $f_P(p) = w_2 \exp(-(x-\mu)^2/e^{-2r})/\sqrt{\pi e^{-2r}}$, which can now be used to obtain the CFI matrix that is diagonal with matrix elements $C_S(1,1) = 2 w_1 e^{-2r}$ and $C_S(2,2) = 2 w_2 e^{-2 r}$. For equal importance of the two estimated parameters the weights can be set to $w_1 = w_2 = 0.5$ and the matrix is identical to the matrix from the entanglement based protocol. This can be also seen from equality $f(x,p) = 2 f_X(x\sqrt{2}) f_P (p\sqrt{2})$. Different importance of the parameters can be then taken into account by adjusting the weights $w_1$ and $w_2$. In all cases, the achieved variances are identical to those obtained in \cite{PhysRevResearch.2.023182} and therefore also saturate the HCR bounds of the entanglement based protocol. Since the energy of the probes that pass through the channel is also identical, the protocol based on separate probes can be also considered optimal.}

\cor{At the level of physical intuition, the equivalence between the schemes can be understood as follows: In the entanglement based scheme, the displacement is encoded simultaneously into both quadratures and entanglement is used}, via the dense-coding effect \cite{Li2002}, to measure them simultaneously. In each run of the experiment, both variables are measured, but there is a cost. The interference on the beam splitter attenuates the displacement so each homodyne detector effectively detects only half of it. On the other hand, \cor{for two sets of squeezed probes}, each individual run detects displacement only in a single quadrature. However, it can detect it fully with no loss of information. \cor{Furthermore, from a practical perspective, the interference in state preparation and detection required by the entanglement based protocol adds to the difficulty of the implementation and causes mode matching losses. For a more thorough analysis on how the losses affect the performance of the protocols, please see the supplementary material. }

\cor{A related problem is the estimation of the amplitude of the displacement $|\alpha| = \sqrt{\mu^2+\nu^2}$ \cite{wolf2019motional, Oh_2020}. This is essentially an estimation of a single parameter that is related to the two parameters of the displacement. It was shown, in \cite{wolf2019motional}, that the optimal estimation strategy with regards to the energy of the probe consists of utilizing photon number Fock states and photon number measurements. For this protocol, the CFI $\mathbf{C}_n = 4\langle a^{\dag}a\rangle +2$ saturates the QFI.}
\cor{Remarkably, even in this case we can approach this performance with the protocol based on the two sets of separately squeezed probes (\ref{2sstate}) with balanced weights $w_1 = w_2 = 0.5$ measured by homodyne detection. The CFI matrix for simultaneous estimation of amplitude $|\alpha|$ and phase $\phi$ can be found to be }
\begin{equation}\label{}
     \mathbf{C} = 
           2(\langle a^{\dag}a\rangle  + 1 + \sqrt{\langle a^{\dag}a\rangle^2 +\langle a^{\dag}a \rangle}) \left(\begin{array}{cc}
           1 & 0 \\
           0 & |\alpha|^2
         \end{array}\right),
\end{equation}
\cor{ where we have used $\langle a^{\dag}a \rangle  = (e^{2r}+e^{-2r}-2)/4$ to allow effective comparison with Fock states and their energy. We can see that in the limit of large energy, the CFI matrix element $\mathbf{C}(1,1)$, corresponding to estimation of $|\alpha|$, approaches the QFI, and thus the ultimate precision, of the single parameter estimation of the Fock based scenario.} For example, with average energy $\langle a^{\dag}a\rangle = 5$ the classical Fisher information for the Fock state approach is $\mathbf{C}_n = 22$, while \cor{for the respective squeezed states with 13 dB squeezing} it is $\mathbf{C}_S = 21.95$. 
At the same time, the approach with independent squeezed states also provides information about the phase of the displacement that is completely disregarded by the Fock state approach.
\cor{The squeezed states are more vulnerable to the adverse effects of loss, because the environment mode vacuum fluctuations always lower bound the achievable variance. See the appendix}. However, their experimental preparation, especially in the optical setting \cite{Takanashi19,Kashiwazaki20,Wollman2015,Pirkkalainen2015,Guarrera2019} is, compared to photon number states and photon number resolving detectors \cite{yukawa2013generating,harder2016local,cooper2013experimental,bohmann2018incomplete}, significantly more feasible.

\cor{
\section{Conclusion}
We have shown that the two parameters of coherent displacement can be independently estimated by two sets of probes prepared in squeezed states and classically correlated position and momentum measurements. The achievable errors are the same as for the optimal methods taking advantage of quantum entanglement of two-mode squeezed states \cite{PhysRevA.87.012107,PhysRevA.97.012106, PhysRevResearch.2.023182} or sub-Planck structure of quantum non-Gaussian probe states \cite{PhysRevA.95.012305}. The equivalence holds asymptotically in the limit of large number of probes, which is a common assumption in quantum estimation scenarios. This finding has several interesting ramifications.
}

\cor{It presents a more feasible scheme for the practical estimation of quantum displacement \cite{mehboudi2015,gilmore2021,PhysRevA.97.032329}, because using sets of differently squeezed states is more feasible and therefore cheaper \cite{Chitambar2019} than the entangled \cite{PhysRevA.87.012107,PhysRevA.97.012106, PhysRevResearch.2.023182} or quantum non-Gaussian states \cite{PhysRevA.95.012305}. We expect it can have a direct impact on sensing of mechanical, electric, magnetic and optical forces \cite{poggio2008,PhysRevA.88.042112,PhysRevA.94.052115}. It also demonstrates that even in multi-parameter quantum estimation it is not necessary to simultaneously estimate all the parameters in each trial, and that mixed states, together with classically correlated measurements, can be optimal. And finally, it also means that the estimation of the two displacement parameters, which is an often studied scenario in quantum metrology, is not a challenging multi-parameter estimation problem, because the two parameters can be efficiently estimated independently. Realizing this and being able to recognize such scenarios will improve our general understanding of quantum multi-parameter estimation protocols and stimulate further theoretical and experimental research. \corb{In the future we hope to analyze the general model arbitrary observables and conclusively identify the scenarios, in which using separable probes is advantageous.}
}

\begin{acknowledgments}
PM acknowledges Project 22-08772S of the Grant Agency of Czech Republic (GACR). KP acknowledges the MEYS of the Czech Republic (Grant Agreement 8C20002) and the funding from the European Union's Horizon 2020 (2014-2020) research and innovation framework program under Grant Agreement No 731473 (ShoQC). Project ShoQC has received funding from the QuantERA ERA-NET Cofund in Quantum Technologies implemented within the European Union's Horizon 2020 Program. RF acknowledges the MEYS under Grant Agreements CZ$.02.1.01/0.0/0.0/16_026/0008460$ and LTAUSA19099. C.O. acknowledges support from the NSF (OMA-1936118). KP, PM, and RF have further been supported by the European Union's 2020 research and innovation programme (CSA - Coordination and support action, H2020-WIDESPREAD-2020-5) under grant agreement No. 951737 (NONGAUSS).

\end{acknowledgments}

\section*{Appendix: Practical comparison with losses}
Let us now expand the model with separate probes to incorporate realistic losses. Since losses are a practical consideration we shall consider only the classical Fisher information that sets the limit of practical measurement tools. Losses in the state preparation manifest by changing the variance matrix of the input state. The variances of the two quadrature operators for a squeezed state need to be represented by two independent values: $V_S$ for the squeezed quadrature and $V_A$ for the anti-squeezed one. In turn, losses in the channel can be modeled as virtual beam splitters which couple the respective mode to a bath in vacuum state. The losses can appear either before or after the estimated operation. These scenarios only differ by a constant scaling factor arising from loss of the displacement. For the sake of comparison between the protocol we can therefore consider losses taking place only before the estimated operation.

For the swapping protocol the model is simple. The initial states are squeezed, with variance $V_S$, and pass through lossy channel with intensity transmission coefficient $\eta_1$. Since the measurement always resolves the squeezed quadrature, the measurement statistics will be again Gaussian function, now with variances
\begin{equation}\label{vout}
    V_{out}  = \eta_1 V_S + (1-\eta_1)\frac{1}{2}.
\end{equation}
This leads to diagonal CFI matrix for the whole measurement that has elements $C_S(1,1) = C_S(2,2) = \frac{2}{V_{out}}.$ We can see that losses in the channel are no fundamentally different from losses in the preparation. In the end, the only thing that matters is the available squeezing.

\begin{figure}
\centering
\includegraphics[width=6.8cm]{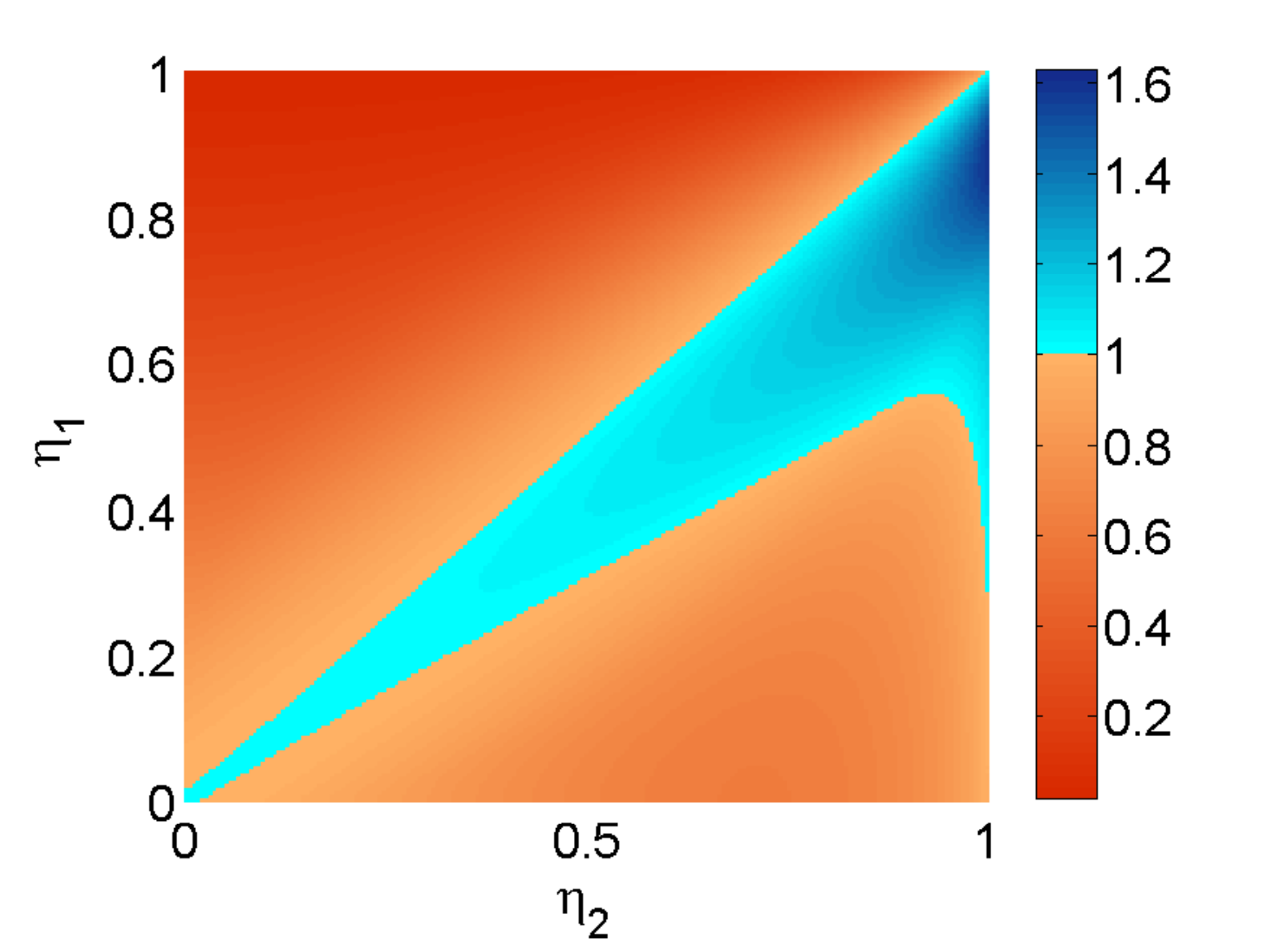}
\caption{Comparing the CFI matrix determinant ratio $\frac{|C_E|}{|C_F|}$ relative to transmission coefficient $\eta_1$ and $\eta_2$ of the two channels. The initial squeezed quantum states are described by $V_S = e^{-2}/2$ and $V_A = e^2/2$. The parameters of the entanglement based scheme were optimized. Blue area - entanglement based protocol has the advantage, Red area - swapping of different probes has the advantage.  }
\label{fig_com}
\end{figure}

The description is more involved in the case the entanglement based protocol. Here it is best to start from the variance matrix of the initial pair of states, $\Sigma = \mathrm{diag}(V_S,V_A,V_A,V_S)$ and use it to find variance matrix of the two modes right before the measurement by subjecting it to sequence of operations:
\begin{equation}\label{sigmaout}
    \Sigma_{out} = O_{BS}^T Y O_{BS} \Sigma_{in} O_{BS}^T Y O_{BS} + \frac{I}{2} - \frac{1}{2}O_{BS}^T Y^2 O_{BS},
\end{equation}
where $O_{BS}$ is the orthogonal matrix describing action of the balanced beam splitter, $I$ is a 4 by 4 unit matrix, and $Y = \mathrm{diag}(\sqrt{\eta_1},\sqrt{\eta_1},\sqrt{\eta_2},\sqrt{\eta_2})$ represents the action of the two lossy channels. The two homodyne detectors, which measure quadratures $x_1$ and $p_2$, then return data with Gaussian distribution with mean values given by the displacement and variance matrix $\Sigma_{m} = \mathrm{diag} (V_m,V_m)$. This matrix is obtained by removing second and third rows and columns from (\ref{sigmaout}) and the variances are
\begin{equation}\label{vm}
    V_m = \frac{1}{4}[ (V_S + V_A-1)(\eta_1+\eta_2) + 2\sqrt{\eta_1 \eta_2} (V_S - V_A) + 2].
\end{equation}
Since the matrix is diagonal with identical diagonal elements, the elements of the diagonal CFI matrix can be obtained as
$    C_E(1,1) = C_E(2,2) = \frac{2}{V_{m}}$
and the comparison between the swapping and the entanglement based protocol fully depends on the two variances (\ref{vout}) and (\ref{vm}).

The first observation that we can make is that when $\eta_1 = \eta_2$, the two protocols are again identical. This is no longer the case when the equality does not hold. From the form of (\ref{vm}) we can see that it describes a parabola for variable $y_2 = \sqrt{\eta_2}$. This parabola has a minimum for
\begin{equation}\label{}
    \sqrt{\eta_2} = \sqrt{\eta_1} \frac{V_A-V_S}{V_A+V_S-1} > \sqrt{\eta_1}
\end{equation}
which means that for any loss $\eta_1$ in the channel 1 containing the channel, there is a range of values $\eta_2<\eta_1$  for loss in the channel 2 for which the protocol has advantage over the swapping scheme which only uses channel 1. The exact range of values of $\eta_2$ depends on the properties of the state. In the limit of large squeezing, in which $V_A$ necessarily approaches infinity, it is optimal to have $\eta_2 = \eta_1$. Interestingly, for the fixed measurement this could lead to the counterintuitive scenario in which it would be beneficial to add artificial losses to the reference arm to achieve optimal regime.

To better understand this behavior we can consider entanglement based protocol that is optimized for the channel transmission coefficient $\eta_1$ and $\eta_2$. The initial quantum states are the same two orthogonally squeezed vacuum states with variances $V_A$ and $V_S$, but the interferometer is now composed of two beam splitters with general coefficients $t_1$, $r_1$ and $t_2$, $r_2$. In such the scenario, the diagonal CFI matrix elements can be found to be
$
C_E(1,1) = \frac{t_2^2}{V_1},\quad C_E(2,2) = \frac{r_2^2}{V_2},
$
with
\begin{align}
V_1 =& 2 t_1 r_1 t_2 r_2 \sqrt{\eta_1 \eta_2}(V_A-V_S) \nonumber \\
&+ \frac{1-t_2^2\eta_1 - r_2^2\eta_2}{2} + \eta_2 r_2^2(t_1^2 V_A + r_1^2 V_S) \nonumber \\
&+ \eta_1 t_2^2(r_1^2 V_A + t_1^2 V_S), \nonumber \\
V_2 =& 2 t_1 r_1 t_2 r_2 \sqrt{\eta_1 \eta_2}(V_A-V_S) \nonumber \\
&+ \frac{1-t_2^2\eta_2 - r_2^2\eta_1}{2} + \eta_1 r_2^2(t_1^2 V_A + r_1^2 V_S) \nonumber \\
&+ \eta_2 t_2^2(r_1^2 V_A + t_1^2 V_S).
\end{align}
We can see that the two diagonal elements are not necessarily equal. This is a consequence of the asymmetrical setup and it separates the approach from the swapping scheme, which is always symmetrical with respect to the two observable quantities. To jointly describe estimation of both quantities we consider the determinant of the CFI matric and numerically maximize it with respect to interferometer parameters $t_1$, $r_1$, $t_2$, and $r_2$. We can than compare the optimal CFI to that of the swapping protocol by means of a ratio
$\frac{|C_E|}{|C_F|}$ that is larger than one when the entanglement protocol has the advantage and smaller than one otherwise. An example of the behavior is shown in Fig.~\ref{fig_com}.

We can see that the swapping protocol has always advantage when $\eta_1 > \eta_2$. However, when $\eta_2 > \eta_1$, the entanglement based protocol has advantage only for some values of $\eta_2$ and those values are close to $\eta_1$ - similarly as without the optimization, there are situations in which it is actually disadvantageous to have losses that are too low. The reason for this is not completely clear but it can follow from the reduced symmetry of the scenario. Nevertheless, with optimization of loss and the interferometer parameters, the entanglement based protocol can always be made to have advantage For $\eta_2 > \eta_1$. This advantage, however, vanishes in the limit of large squeezing.

\begin{figure}
\centering
\includegraphics[width=6.8cm]{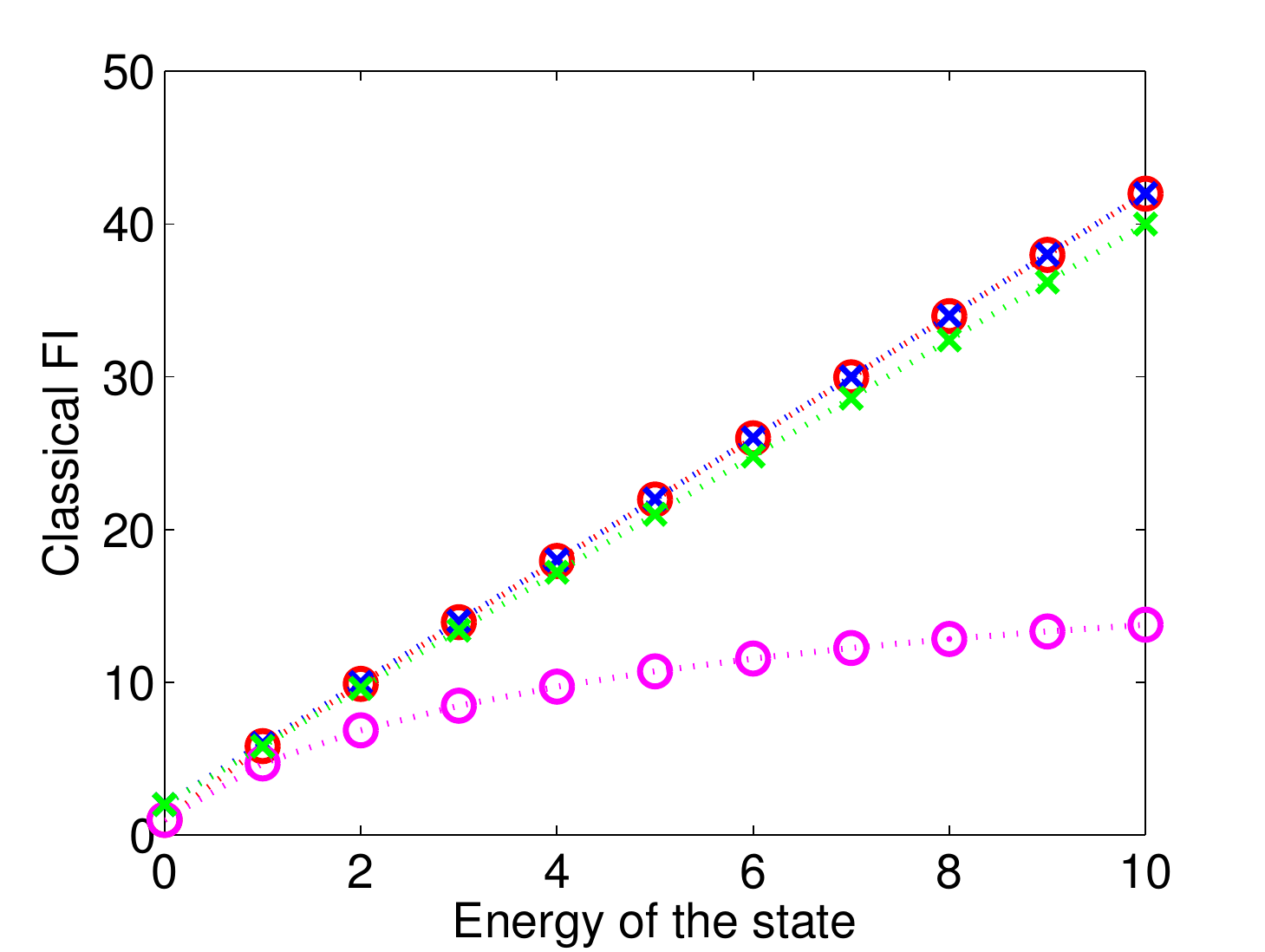}
\caption{Comparison of classical Fisher information for estimation of amplitude of displacement with lossy states. Blue crosses - Fock state approach with $\eta = 1$. Green crosses - Fock state approach with $\eta = 0.95$. Red circles - squeezed state approach with $\eta = 1$. Magenta circles - squeezed state approach with $\eta = 0.95$.}
\label{Fig_loss}
\end{figure}

In estimation of the amplitude of the displacement with no regards the phase, we can again use (\ref{vout}) and, under the assumption the states were pure with variance $V_S = e^{-2r}/2$ before the channel, find the relevant classical Fisher information matrix element equal to
\begin{equation}\label{}
    C_S(1,1) = \frac{1}{2 V_{out}}
\end{equation}
where the initial energy of the state is given by $\langle a^{\dag}a\rangle = \frac{e^{2r} + e^{-2r}-2}{4}$. For the scheme employing the Fock states we can evaluate the quantum Fisher information because it is saturated by the CFI. The quantum FI for Fock state $|n\rangle$ can be now found to be
\begin{align}\label{}
    Q_F = 2 \sum_{k = 0}^{n} \left( \begin{array}{c}
                                   n \\
                                   k
                                 \end{array} \right) \eta^{k}(1-\eta)^{n-k}(2k+1).
\end{align}
the comparison of the classical Fisher information for the two scenarios and for $\eta=1$ and $\eta = 0.95$ is shown in Fig.~\ref{Fig_loss}.
We can see that in the ideal scenario with $\eta = 1$ the two approaches are practically identical. However, while the losses only marginally affect the Fock state scenario, they significantly reduce the performance of the squeezed state protocol. This is the consequence of the form of (\ref{vout}) that is lower bounded by the losses no matter what is the initial energy of the state.

%

\end{document}